\documentclass[12pt]{iopart}

\usepackage{graphicx}
\usepackage{amssymb}

\begin{document}

\title{Entanglement in pure and thermal cluster states}

\author{Michal Hajdu\v{s}ek$^1$ and Vlatko Vedral$^{2,3,4}$}

\address{$^1$The School of Physics and Astronomy, University of
Leeds, Leeds, LS2 9JT, United Kingdom}
\address{$^2$Centre for Quantum
Technologies, National University of Singapore, 3 Science Drive 2,
Singapore, 117543}
\address{$^3$Department of Physics, National
University of Singapore, 3 Science Drive 2, Singapore, 117543}
\address{$^4$Atomic and Laser Physics, Clarendon Laboratory, University of Oxford, Parks Road, Oxford OX1 3PU, United Kingdom}

\ead{phymha@leeds.ac.uk}

\begin{abstract}
We present a closest separable state to cluster states. We start by
considering linear cluster chains and extend our method to cluster
states that can be used as a universal resource in quantum
computation. We reproduce known results for pure cluster states and
show how our method can be used in quantifying entanglement in noisy
cluster states. Operational meaning is given to our method that
clearly demonstrates how these closest separable states can be
constructed from two-qubit clusters in the case of pure states. We
also discuss the issue of finding the critical temperature at which
the cluster state becomes only classically correlated and the
importance of this temperature to our method.
\end{abstract}


\maketitle

\section{Introduction}
Entanglement plays an important role in modern physics and is the
subject of intense research both for its implications in our
fundamental understanding of nature \cite{EPR:1935} as well as its
practical usefulness in quantum information processing
\cite{Walther:2005} and \cite{Vandersypen:2001}. Different models of
quantum computing have been devised to harness this usefulness. The
most prominent ones include the circuit model \cite{Deutsch:1985},
topological computing \cite{Freedman:2003} and measurement-based
computing \cite{Raussendorf:2001}.

Measurement-based quantum computation (or one-way quantum
computation) differs from the circuit model by using an initially
prepared highly entangled resource state. The algorithm proceeds as
a set of local measurements on this state. The basis and order of
these measurements characterize the algorithm and outcomes of
previous measurements are fed forward to make the computation
deterministic.

The general resource for this computation model are graph states
\cite{Hein:2004}. In this paper we will concentrate on particular
graph states, namely cluster states \cite{Briegel:2000}. Unlike
graph states, cluster states have regular square structure. One of
the advantages of one-way quantum computation is that all the
entanglement used in the algorithm is present right at the
beginning. The subsequent measurements only consume it. This makes
it easier to identify the role of entanglement in the information
processing compared to the traditional circuit model. Recently it
has also been shown both theoretically and experimentally in
\cite{Biggerstaff:2009} that the quantum computational power of
cluster states can be extended by replacing some projective
measurements with generalized quantum measurements. This makes
cluster state quantum computation a very attractive and promising
prospect when it comes to constructing a quantum computer.

Due to its clear role as a resource in one-way quantum computation
it is highly desirable to know the entanglement properties of
cluster states. Perhaps one of the most important problems in
studying the properties of quantum states, and resource states for
quantum computation in particular, is the quantification of
entanglement itself. The entanglement scaling of pure
$d$-dimensional graph states has been shown by Markham et al to be
$N/2$ \cite{Markham:2007}. However any realistic quantum computer
will operate at finite temperatures which makes it necessary to be
able to characterize entanglement of mixed states. Quantifying
entanglement is a notoriously difficult task \cite{Horodecki:2009}
and the exact scaling of entanglement in thermal cluster states is
still unknown though some advances have been made by investigating
the localization of entanglement in noisy cluster states
\cite{Raussendorf:2005}.

There are many entanglement measures that quantify the scaling of
entanglement in quantum states. In this work we use relative entropy
of entanglement (see \cite{Vedral:1997} and \cite{Vedral:1998}) to
quantify the scaling. In general there are not many multipartite
quantum states for which the relative entropy of entnaglement can be
calculated analytically. $D$-dimensional symmetric states are one of
the few examples \cite{Wei:2004}, \cite{Vedral:2004}. It is unknown
how to find the closest separable state to a general entangled state
and the problem of closed form of relative entropy still remains an
open question. Though some progress has been made in this direction
(\cite{Ishizaka:2003} and \cite{Miranowicz:2008}) it is impossible
to do this even for a general two-qubit state.

We present a method of constructing the closest separable state to a
thermal cluster and show how it can reproduce the results of
\cite{Markham:2007} for pure cluster states. We also show how bound
entanglement complicates the procedure of finding the closest
separable state for noisy cluster states.

\section{Closest separable state for pure cluster states\label{pure}}
The entanglement of cluster states is known to scale linearly with
the size of the system as $\frac{N}{2}$. This result was first
obtained by Markham et al in \cite{Markham:2007}. The authors used a
technique where they derived upper and lower bounds on the
entanglement and then they showed that these two bounds are equal.

In this section we verify this result using geometrical ideas. To do
this we employ the relative entropy of entanglement
(\cite{Vedral:1997},\cite{Vedral:1998}) $E(\sigma)$ defined as
\begin{equation} \label{eq:relative_entropy}
E(\sigma)=\min_{\rho\in\mathfrak{D}}S(\sigma||\rho)
\end{equation}
where $\sigma$ is the entangled state. The minimization is taken
over the set $\mathfrak{D}$ of all separable states $\rho$ and
$S(\sigma||\rho)=\textrm{Tr}[\sigma\log\sigma-\sigma\log\rho]$ is
the relative entropy. For pure states this simplifies to
$S(\sigma||\rho)=-\textrm{Tr}[\sigma\log\rho]$. Since we are
considering only qubits we take the logarithm to base 2. State that
minimizes the relative entropy in (\ref{eq:relative_entropy}) will
be designated by $\rho^{\ast}$ and will be called the closest
separable state. In order to prove that this state $\rho^{\ast}$
achieves the minimum of relative entropy we employ the same approach
as \cite{Vedral_thesis:1998}. Consider a small region around the
closest separable state $(1-x)\rho^{\ast}+x\tau$ for a small
parameter $x$. All we have to show now is that for a general
separable state $\tau$ the gradient of relative entropy is
non-negative in this region. Due to convexity of the set of
separable states this will also mean that the minimum of relative
entropy is global.

A cluster state can be thought of as a state whose qubits are first
prepared in the +1 eigenstate of $\sigma^{x}$ and then control-phase
gates $CZ$ are applied between nearest neighbours. For example the
two-qubit cluster state is
$|\psi_{2}\rangle=CZ|+\rangle|+\rangle=\frac{1}{2}[|00\rangle+|01\rangle+|10\rangle-|11\rangle]$.
It can be easily seen that the closest separable state
$\rho_{2}^{\ast}$ is
\begin{equation} \label{eq:2_closest}
\rho_{2}^{\ast}= \frac{1}{4}\left( \begin{array}{ccrr} 1 & 1 & 0 & 0 \\
1 & 1 & 0 & 0\\
0 & 0 & 1 & -1\\
0 & 0 & -1 & 1
\end{array} \right)
\end{equation}
by calculating that $E(\sigma_{2}||\rho_{2}^{\ast})=1$ where
$\sigma_{2}=|\psi_{2}\rangle\langle\psi_{2}|$.

So far all the states have been expressed in the computational basis
$\{|0\rangle,|1\rangle\}$. It is however more useful to write the
states in what we call the "mixed" basis
$\{|0\rangle,|1\rangle,|+\rangle,|-\rangle\}$. So the state vector
and the closest separable state for the case of 2 qubits now become
\begin{eqnarray}\label{eq:2_cluster_and_sep}
|\psi_{2}\rangle & = & \frac{1}{\sqrt{2}}[|0+\rangle+|1-\rangle]\\
\rho_{2}^{\ast} & = &
\frac{1}{2}[|0+\rangle\langle0+|+|1-\rangle\langle1-|]
\end{eqnarray}
Similarly for 4 qubits the expressions take the following form
\begin{eqnarray}\label{eq:4_cluster_and_sep}
|\psi_{4}\rangle & = &
\frac{1}{2}[|0+0+\rangle+|0-1-\rangle+|1-0+\rangle+|1+1-\rangle]\\
\rho_{4}^{\ast} & = &
\frac{1}{4}[|0+0+\rangle\langle0+0+|+|0-1-\rangle\langle0-1-|\\\nonumber
&&+|1-0+\rangle\langle1-0+|+|1+1-\rangle\langle1+1-|]
\end{eqnarray}
From the form of $\rho_{2}^{\ast}$ and $\rho_{4}^{\ast}$ we can see
that the closest separable state is obtainable from the state vector
by writing the density matrix in the "mixed" basis and then keeping
only the diagonal terms. This is why it is more instructive to work
in the "mixed" basis, one can immediately see the form of the
closest separable state from the state vector. Of course in the
computational basis not all off-diagonal terms are zero.

Important thing to notice is the number of terms in the state vector
of a cluster state is different in the computational basis and in
the "mixed" basis. In the computational basis all the coefficients
are non-zero, they are either $1$ or $-1$. So for $N$-qubit cluster
state there are $2^{N}$ basis coefficients. On the other hand in the
"mixed" basis the number of coefficients drops to $2^{N/2}$. So when
it comes to normalization coefficients for the density matrices we
have

\begin{center}
\begin{tabular}{|l | c | c | c|}
\hline & Vector $|\psi_{N}\rangle$ & Cluster $\sigma_{N}$ &
Separable $\rho_{N}^{\ast}$\\ \hline Computational basis &
$2^{-N/2}$ & $2^{-N}$ & $2^{-N}$\\ \hline
"Mixed" basis & $2^{-N/4}$ & $2^{-N/2}$ & $2^{-N/2}$\\
\hline
\end{tabular}
\end{center}
From the above it is clear to see that the closest separable matrix
$\rho_{N}^{\ast}$ is of rank $2^{N/2}$. Another useful fact that
will be used later is that when the matrices are expressed in the
"mixed" basis it is easy to see that they commute with the cluster
state $[\rho_{N}^{\ast},\sigma_{N}]=0$.

Before we start the proof for the closest separable state a small
note about our notation is in place. The proof contains places where
in stead of just writing $\sigma_{N}$ or $\rho_{N}^{\ast}$ it is
more instructive to write out the state explicitly, for instance for
2 qubits
$\sigma=\frac{1}{2}(|0+\rangle+|1-\rangle)*(\langle0+|+\langle1-|)$.
This is of course not possible in the case of general $N$. Therefore
we use the following notation.
\begin{eqnarray}
\textrm{Pure
vector}\qquad\qquad|\psi_{N}\rangle=2^{-N/4}(|...\rangle+...)\nonumber\\
\textrm{Pure
matrix}\qquad\qquad\sigma_{N}=2^{-N/2}(|...\rangle+...)*(\langle...|+...)\nonumber\\
\textrm{Closest sep.
state}\qquad\rho_{N}^{\ast}=2^{-N/2}(|...\rangle\langle...|+...)\nonumber
\end{eqnarray}
From the normalization factors it is clear that the states are in
the "mixed" basis. This also implies that each round bracket
contains $2^{N/2}$ terms.

As mentioned above, to prove that states of the form
$\rho_{2}^{\ast}$ and $\rho_{4}^{\ast}$ are really the closest
separable states we need to show that the gradient of relative
entropy in a small region around $\rho_{N}^{\ast}$ is non-negative
\begin{equation} \label{eq:N_gradient_S}
\lim_{x\rightarrow0}\frac{\partial}{\partial
x}S(\sigma||(1-x)\rho_{N}^{\ast}+x\tau )\geq0
\end{equation}
Substituting $f(x,\tau)=S(\sigma||(1-x)\rho_{N}^{\ast}+x\tau)$ and
realizing that for a positive operator $A$ we have the expression
$\log A=\int_{0}^{\infty}\frac{At-1}{A+t}\frac{dt}{1+t^{2}}$
condition (\ref{eq:N_gradient_S}) becomes
\begin{equation}\label{eq:N_gradient_f}
\frac{\partial f}{\partial
x}(0,\tau)=\textrm{Tr}[\sigma\int_{0}^{\infty}(\rho_{N}^{\ast}+t)^{-1}(\rho_{N}^{\ast}-\tau)(\rho_{N}^{\ast}+t)^{-1}dt]\geq
0
\end{equation}
Starting with the first integral in Eq.(\ref{eq:N_gradient_f})
\begin{eqnarray}\label{eq:first_int}
&=&\textrm{Tr}[\sigma\int_{0}^{\infty}(\rho_{N}^{\ast}+t)^{-1}\rho_{N}^{\ast}(\rho_{N}^{\ast}+t)^{-1}dt]\nonumber\\
&=&\textrm{Tr}[\sigma\int_{0}^{\infty}(\rho_{N}^{\ast}+t)^{-2}dt\rho_{N}^{\ast}]
\end{eqnarray}
Evaluating the integral in Eq.(\ref{eq:first_int}):
\begin{eqnarray}
\int_{0}^{\infty}(\rho_{N}^{\ast}+t)^{-2}dt&=&\int_{0}^{\infty}(2^{-N/2}+t)^{-2}dt(|...\rangle\langle...|+...)\nonumber\\
&=&\int_{0}^{\infty}(2^{-N/2}+t)^{-2}dt*2^{N/2}\rho_{N}^{\ast}\nonumber\\
&=&2^{N/2}*2^{N/2}\rho_{N}^{\ast}\nonumber\\
&=&2^{N}\rho_{N}^{\ast}
\end{eqnarray}
Substituting this back into Eq.(\ref{eq:first_int}):
\begin{eqnarray}
\textrm{Tr}[\sigma*2^{N}\rho_{N}^{\ast}\rho_{N}^{\ast}]&=&2^{N}[\sigma*2^{-N}(|...\rangle\langle...|+...)*(|...\rangle\langle...|+...)]\\
&=&\textrm{Tr}[\sigma(|...\rangle\langle...|+...)]\nonumber\\
&=&2^{-N/2}*2^{N/2}=1\nonumber
\end{eqnarray}
Now turning our attention to the second integral in
Eq.(\ref{eq:N_gradient_f}). Here we will use the fact that $\sigma$
and $\rho_{N}^{\ast}$ commute with each other.
\begin{eqnarray}
&=&\textrm{Tr}[\int_{0}^{\infty}(\rho_{N}^{\ast}+t)^{-1}\sigma(\rho_{N}^{\ast}+t)^{-1}dt\tau]\nonumber\\
&=&\textrm{Tr}[\int_{0}^{\infty}(\rho_{N}^{\ast}+t)^{-2}dt\sigma\tau]\nonumber\\
&=&2^{N}\textrm{Tr}[\rho_{N}^{\ast}\sigma\tau]\nonumber\\
&=&2^{N}\textrm{Tr}[2^{-N/2}2^{-N/2}(|...\rangle\langle...|+...)*(|...\rangle+...)*(\langle...|+...)\tau]\nonumber\\
&=&\textrm{Tr}[(|...\rangle\langle...|+...)*(|...\rangle+...)*(\langle...|+...)\tau]\nonumber\\
&=&\textrm{Tr}[(|...\rangle+...)*(\langle...|+...)\tau]\nonumber\\
&=&2^{N/2}\textrm[\sigma\tau]
\end{eqnarray}
So finally the gradient of the relative entropy in
Eq.(\ref{eq:N_gradient_f}) can be written as
\begin{equation}\label{eq:end}
\frac{\partial f}{\partial
x}(0,\tau)=1-2^{N/2}\textrm{Tr}[\sigma\tau]
\end{equation}

\begin{figure}[t]
\begin{center}
\includegraphics[scale=0.6]{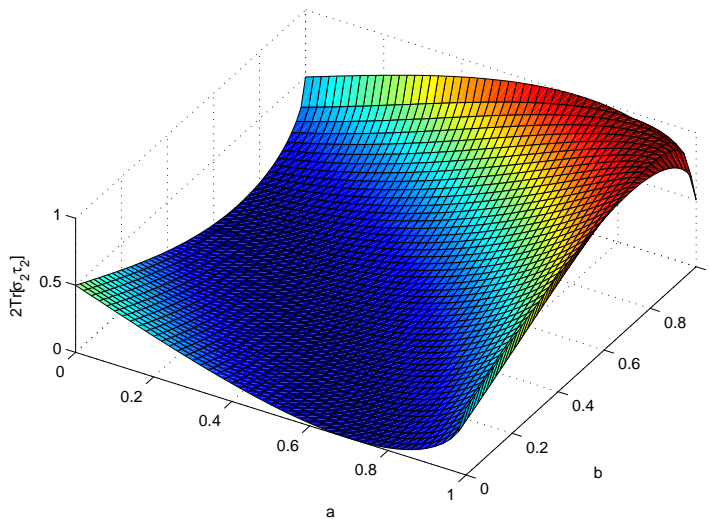}
\end{center}
\caption{Plot of $2\textrm{Tr}[\sigma_{2}\tau_{2}]$ showing that the
maximum of the trace is 1 and therefore the gradient of relative
entropy is non-negative. Here $a$ and $b$ are parameters from $\tau$
representing the different amplitudes for the two subsystems. So
$\tau=|\alpha\beta\rangle\langle\alpha\beta|$ where
$|\alpha\rangle=a|0\rangle+\sqrt{1-|a|^{2}}|1\rangle$ and similarly
$|\beta\rangle=b|0\rangle+\sqrt{1-|b|^{2}}|1\rangle$.}\label{fig:2_gradient}
\end{figure}

In order for inequality (\ref{eq:end}) to be satisfied the trace
needs to scale at most as $2^{-N/2}$. This is in fact the case as
can be proved by induction. First we will consider
$\tau=|\alpha\beta\rangle\langle\alpha\beta|$. The inequality has
been verified numerically for 2, 4 and 6 qubits. In the case of 2
qubits graphical solution has been obtained as well in
Fig.(\ref{fig:2_gradient}).
\begin{eqnarray}\label{eq:2_4_6_trace}
N=2\qquad\textrm{Tr}[\sigma_{2}\tau_{2}]\leq\frac{1}{2}=2^{-1}\\
N=4\qquad\textrm{Tr}[\sigma_{4}\tau_{4}]\leq\frac{1}{4}=2^{-2}\nonumber\\
N=6\qquad\textrm{Tr}[\sigma_{6}\tau_{6}]\leq\frac{1}{8}=2^{-3}\nonumber
\end{eqnarray}
Assuming that the above holds true for the case of $N=k$, namely
$\textrm{Tr}[\sigma_{k}\tau_{k}]\leq2^{-k/2}$. Now we need to prove
using the above assumption that the case $N=k+2$ holds true
\begin{equation}
\textrm{Tr}[\sigma_{k+2}\tau_{k+2}]\leq2^{-k/2-1}
\end{equation}
Starting with the trace
\begin{eqnarray}\label{eq:trace_proof}
\textrm{Tr}[\sigma_{k+2}\tau_{k+2}]&=&\langle\psi_{k+2}|\tau_{k+2}|\psi_{k+2}\rangle\\
&=&\langle\psi_{k}|\langle\psi_{2}|CZ^{\dagger}\tau_{k}\otimes\tau_{2}CZ|\psi_{k}\rangle|\psi_{2}\rangle\nonumber
\end{eqnarray}
where we have used the fact that
$|\psi_{k+2}\rangle=CZ|\psi_{k}\rangle|\psi_{2}\rangle$ and
$\tau_{k+2}=\tau_{k}\otimes\tau_{2}$. The control-phase gate acts on
qubits $k$ and $k+1$. Also the separable state $\tau$ does not
commute with the controlled phase gate so their commutator is
non-zero $[\tau_{k}\otimes\tau_{2},CZ]=A$ so we can substitute
$\tau_{k}\otimes\tau_{2}CZ=A+CZ\tau_{k}\otimes\tau_{2}$ into
Eq.(\ref{eq:trace_proof}).
\begin{eqnarray}
\textrm{Tr}[\sigma_{k+2}\tau_{k+2}]&=&\langle\psi_{k}|\langle\psi_{2}|CZ^{\dagger}A|\psi_{k}\rangle|\psi_{2}\rangle\\
&+&\langle\psi_{k}|\tau_{k}|\psi_{k}\rangle*\langle\psi_{2}|\tau_{2}|\psi_{2}\rangle\nonumber
\end{eqnarray}
Taking a closer look at the product $CZ^{\dagger}A$ and using the
above commutation relation we get
$CZ^{\dagger}A=CZ^{\dagger}\tau_{k}\otimes\tau_{2}CZ-\tau_{k}\otimes\tau_{2}$.
Taking the trace of both sides
\begin{equation}
\textrm{Tr}[CZ^{\dagger}A]=\textrm{Tr}[CZ^{\dagger}\tau_{k}\otimes\tau_{2}CZ]-\textrm{Tr}[\tau_{k}\otimes\tau_{2}]=0
\end{equation}
Using the assumption for $N=k$ this implies
\begin{equation}
\textrm{Tr}[\sigma_{k+2}\tau_{k+2}]=\langle\psi_{k}|\tau_{k}|\psi_{k}\rangle*\langle\psi_{2}|\tau_{2}|\psi_{2}\rangle\leq2^{-k/2-1}
\end{equation}
Therefore $(\partial f/\partial
x)(0,|\alpha\beta\ldots\rangle\langle\alpha\beta\ldots|)\geq0$.
Since any separable state can be written in the form
$\rho=\sum_{i}p^{i}|\alpha^{i}\beta^{i}\ldots\rangle\langle\alpha^{i}\beta^{i}\ldots|$
we have
\begin{equation}
\frac{\partial f}{\partial x}(0,\rho)=\sum_{i}p^{i}\frac{\partial
f}{\partial
x}(0,|\alpha^{i}\beta^{i}\ldots\rangle\langle\alpha^{i}\beta^{i}\ldots|)\geq0
\end{equation}

Therefore $\rho_{N}^{\ast}$ is the closest separable state to
$\sigma_{N}$. Now we can compute the relative entropy between the
cluster state $\sigma_{N}$ and $\rho_{N}^{\ast}$.
\begin{eqnarray}
E(\sigma_{N})&=&S(\sigma_{N}||\rho_{N}^{\ast})\\
&=&-\textrm{Tr}[\sigma_{N}\log\rho_{N}^{\ast}]\nonumber\\
&=&-\langle\psi_{N}|\log\rho_{N}^{\ast}|\psi_{N}\rangle\nonumber\\
&=&-\log\langle\psi_{N}|\rho_{N}^{\ast}|\psi_{N}\rangle\nonumber\\
&=&-\log2^{-N/2}2^{-N/2}(\langle...|+...)*(|...\rangle\langle...|+...+|...)*(|...\rangle+...+|...)\nonumber\\
&=&-\log2^{-N}2^{N/2}\nonumber\\
&=&-\log2^{-N/2}=\frac{N}{2}
\end{eqnarray}
This concludes the proof that our closest separable state gives
linear scaling of entanglement and reproduces the results of
\cite{Markham:2007}.

\begin{figure}
\begin{center}
\includegraphics[scale=0.8]{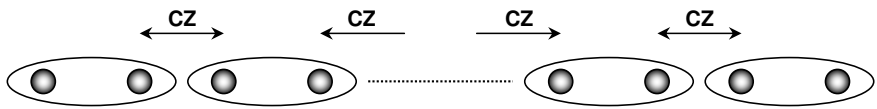}
\end{center}
\caption{Closest separable state of $N$ qubits. Each pair of qubits
represents the closest separable state
$\rho_{2}^{\ast}$.}\label{fig:N_separable_CZ}
\end{figure}

We can look at the form of the closest separable state from a more
operational point of view. Imagine we are asked to prepare the
closest separable state to an $N$-qubit one-dimensional cluster
state. The only ingredients needed to construct this state are the
2-qubit closest separable state $\rho_{2}^{\ast}$ and control-phase
gates $CZ$. The state $\rho_{N}^{\ast}$ can be written
\begin{equation}\label{eq:N_closest_sep}
\rho_{N}^{\ast}=U*\underbrace{(\rho_{2}^{\ast}\otimes\rho_{2}^{\ast}\otimes\ldots\otimes\rho_{2}^{\ast})}_{N/2}*U^{\dagger}
\end{equation}
where the unitary applied has the form
\begin{equation}\label{eq:unitary_CZ}
U=I\otimes \underbrace{CZ\otimes CZ\otimes\ldots\otimes
CZ}_{N/2-1}\otimes I
\end{equation}
All the unitary (\ref{eq:unitary_CZ}) implements are controlled
phase gates $CZ$ between qubits 2-3 and 4-5 and so on. It connects
the states $\rho_{2}^{\ast}$ together in a chain as shown in
Fig.(\ref{fig:N_separable_CZ}).
It is important to notice that although control-phase gates are
entangling operations, they leave the state $\rho_{N}^{\ast}$ in a
separable form.

This can easily be seen if we consider two separable 2-qubit states
(\ref{eq:2_closest}) joined by a control-phase operation. The total
4-qubit state can be written as
$\rho_{4}^{\ast}=CZ_{2,3}\rho_{2}^{\ast}\otimes\rho_{2}^{\ast}CZ_{2,3}^{\dag}$.
This state is locally equivalent to
$\rho_{4}^{\ast}=\frac{1}{4}[|0000\rangle\langle0000|+|0111\rangle\langle0111|+|1011\rangle\langle1011|+|1100\rangle\langle1100|]$
where we have applied Hadamard operators on every second qubit
$I\otimes H\otimes I\otimes H$. The transformed state
$\rho_{4}^{\ast}$ is clearly separable. Since identical analysis can
be applied to $\rho_{N}^{\ast}$ we conclude that control-phase
operations in the unitary (\ref{eq:unitary_CZ}) do not introduce any
entanglement and $\rho_{N}^{\ast}$ remains separable.

It is important to note that the form of the closest separable state
is the same for cluster states in higher dimensions as well. The
majority of the proof for closest separable states is independent of
the dimension of our cluster state. The dimensionality becomes
important only in Eq.(\ref{eq:trace_proof}) where we assume
one-dimensional cluster state by using
$|\psi_{k+2}\rangle=CZ|\psi_{k}\rangle|\psi_{2}\rangle$ where the
control-phase gate is applied between qubits $k$ and $k+1$. For
higher dimensions the number of control-phase gates in
Eq.(\ref{eq:trace_proof}) increases due to larger number of
neighbouring qubits. However the logic of the proof remains
unchanged. Therefore it is possible to find the closest separable
state to a pure cluster state of any dimensionality. This is crucial
because unlike linear cluster chains, two and higher dimensional
clusters are universal resources for quantum computation
\cite{Raussendorf:2001}.

\section{Thermal entanglement\label{thermal_ent}}
Study of thermal entanglement is of great significance since any
realistic scheme of implementing a quantum computer will operate at
finite  temperatures. In this section we look at entanglement in
thermal cluster states. Specifically we are interested to see at
what temperature the cluster state becomes separable. These
considerations will lead us to a more general form for the closest
separable state.

Consider the cluster Hamiltonian
\begin{equation}\label{eq:stabiliser_hamiltonian}
H=-J\sum_{j=1}^{N}\sigma_{j}^{x}\bigotimes_{i\in\mathcal{N}}\sigma_{i}^{z}
\end{equation}
where $J$ is the coupling constant and $\mathcal{N}$ is the
neighbourhood of site $j$. The terms in the sum above are just a
particular case of stabilizer operators \cite{Raussendorf:2001}.
Hence Hamiltonian (\ref{eq:stabiliser_hamiltonian}) is also referred
to as the stabilizer Hamiltonian. The ground state of this
Hamiltonian is the cluster state. Excited states are achieved by
local $\sigma_{j}^{z}$ flips. The ground state has energy $-NJ$ and
is non-degenerate. The $k^{th}$ excited state has energy $J(-N+2k)$
with degeneracy $\frac{N!}{k!(N-k)!}$ \cite{Markham:2008}. All the
states in the spectrum of the Hamiltonian are equally entangled
since they are all some $\sigma^{z}$ away from the ground state.
Therefore our method of finding the closest separable state can also
be used on these excited states.

The ground state of this Hamiltonian is highly entangled and the
entanglement scales linearly \cite{Markham:2007} with the size of
the cluster state as $\frac{N}{2}$. An interesting question to ask
is whether this entanglement persists at finite temperatures and
where is the critical point beyond which the thermal entangled state
becomes separable. Thermal cluster state has the following form
\cite{Raussendorf:2005}
\begin{equation}\label{eq:thermal_cluster}
\sigma(\beta)=\frac{1}{2^{N}}\prod^{N}_{j}[I+\tanh(\beta J)K_{j}]
\end{equation}
where $K_{j}$ is the stabilizer operator at site $j$. For simplicity
we will now consider a two-qubit thermal cluster state of the form
\begin{equation}\label{eq:2_thermal_cluster}
\sigma_{2}(\omega)=\frac{1}{4}(I+\omega\sigma^{x}\otimes\sigma^{z})(I+\omega\sigma^{z}\otimes\sigma^{x})
\end{equation}
where $\omega=\tanh(\beta J)$. So in the low temperature limit as
$T\rightarrow0$ we recover the pure state because
$\omega\rightarrow1$. As the temperature increases
$\omega\rightarrow0$ and for $T\rightarrow\infty$ we obtain a
maximally mixed state. To calculate the critical temperature at
which the state becomes separable we use the Peres-Horodecki
criterion \cite{Peres:1995} and \cite{Horodecki:1996}. To see when
the state fails to be a positive definite we have to solve the
following equation $\omega_{c}^{2}+2\omega_{c}-1=0$. This equation
has two solutions $\omega_{c}=-1\pm\sqrt{2}$ but we will disregard
the negative solution because it is not physical. Substituting this
back into $\omega=\tanh(\beta J)$ finally gives us the critical
temperature
\begin{equation}\label{eq:crit_temp}
T_{c}=-\frac{2J}{k_{B}\ln(\sqrt{2}-1)}
\end{equation}
The entanglement for $T<T_{c}$ happens to be of useful distillable
nature \cite{Kay:2006}. Therefore by using local operations and
classical communication on multiple copies of the thermal cluster
pure entanglement can be distilled. This critical temperature for
distillable entanglement remains unchanged for higher dimensions and
any system size $N$ as proved in \cite{Markham:2008}. For
temperatures above the critical temperature, $T\geq T_{c}$, the
$N$-qubit cluster state, where $N>2$, is not fully separable but is
bound entangled \cite{Cavalcanti:2009}. The critical temperature at
which all entanglement vanishes is non-trivial to find due to the
fact that properties of bound entanglement are still not fully
understood.

We have already calculated the closest separable state for a 2-qubit
cluster. Now we will show that this state is not unique. In fact
there are infinitely many states that satisfy condition
(\ref{eq:N_gradient_S}). Since there is no bound entanglement for
the case of two qubits we know that the thermal cluster state at the
critical temperature $T_{c}$ is fully separable. It is natural to
ask the question what the distance is between this state and a pure
cluster state in terms of the relative entropy. So we want to
calculate $S(\sigma_{2}||\sigma_{2}(\omega_{c}))$ where
$\sigma_{2}(\omega_{c})=\frac{1}{4}(I+\omega_{c}\sigma^{x}\otimes\sigma^{z})(I+\omega_{c}\sigma^{z}\otimes\sigma^{x})$
and $\omega_{c}=\sqrt{2}-1$. It can be quickly checked that
$S(\sigma||\sigma_{2}(\omega_{c}))=1$ and the gradient of relative
entropy in a small region around $\sigma_{2}(\omega_{c})$ is
$\frac{\partial f}{\partial x}(0,\tau)=1-2\textrm{Tr}[\sigma\tau]>0$
as can be seen from Eq.(\ref{eq:2_4_6_trace}). Therefore the 2-qubit
thermal cluster at critical temperature is another good closest
separable state to the pure cluster state.

It is instructive to look at the thermal cluster state
$\sigma_{2}(\omega)$ in the "mixed" basis and compare it to the pure
cluster closest separable state $\rho_{2}^{\ast}$. Transforming
$\sigma_{2}(\omega)$ we have
\begin{equation}\label{eq:transf}
(I\otimes H)\sigma_{2}(\omega)(I\otimes H)^{\dag}=\frac{1}{4}\left( \begin{array}{ccrr} 1+\omega & 0 & 0 & \omega(1+\omega) \\
0 & 1-\omega & \omega(\omega-1) & 0\\
0 & \omega(\omega-1) & 1-\omega & 0\\
\omega(1+\omega) & 0 & 0 & 1+\omega
\end{array} \right)
\end{equation}
From the form of the matrix it can be seen that the only way that
$\sigma_{2}(\omega)\rightarrow\rho_{2}^{\ast}$ is when all the terms
apart from $1+\omega$ vanish. So we arrive at the following four
conditions that have to be satisfied simultaneously;
$\omega(1+\omega)=1-\omega=\omega(\omega-1)=0$ and $1+\omega=2$. It
is straightforward to see that the only solution is $\omega=1$ which
corresponds to the pure cluster state case. Therefore as the
temperature increases the thermal cluster state does not approach
the pure closest separable state $\rho_{2}^{\ast}$. This last result
can be seen also from the fact that the relative entropy between
these two states is non-zero, ie.
$S(\rho_{2}^{\ast}||\sigma_{2}(\omega_{c}))\neq0$. The relative
entropy between two states $S(A||B)$ vanishes if and only if $A=B$
\cite{Barnett:2009}.

Taking a convex mixture of $\rho_{2}^{\ast}$ and
$\sigma_{2}(\omega_{c})$ allows us to find a general form of the
closest 2-qubit separable state
\begin{equation}\label{eq:2_closest_general}
\rho_{2}=(1-\lambda)\rho_{2}^{\ast}+\lambda\sigma_{2}(\omega_{c})
\end{equation}
where $\lambda\in[0,1]$. This can be verified by computing the
gradient of the relative entropy in the usual way, $\frac{\partial
f}{\partial x}(0,\tau)=1-2\textrm{Tr}[\sigma\tau]\geq0$ for all
$\lambda\in[0,1]$.

\section{Thermal cluster states\label{therm_clust}}
Now we are in a position to calculate the entanglement of thermal
cluster states. We will use a similar approach as above to show that
the form of the closest separable state is the same as in
Eq.(\ref{eq:2_closest_general}). However this time the parameter
$\lambda$ that determines the mixture of our two closest separable
states from Eq.(\ref{eq:2_closest_general}) will not be completely
free and its lower bound will depend on the temperature. We will
show that as the temperature increases towards the critical
temperature $T\rightarrow T_{c}$ the parameter $\lambda$ increases
towards unity.

\begin{figure}
\begin{center}
\includegraphics[scale=0.7]{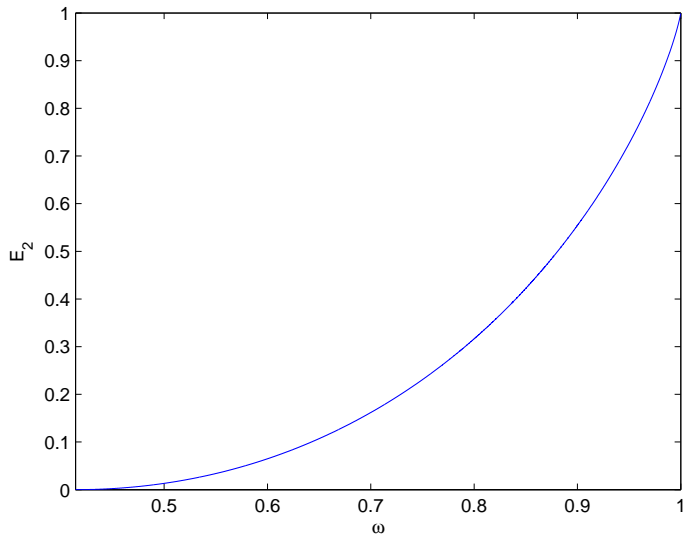}
\end{center}
\caption{Scaling of entanglement of 2 and 4 qubits with increasing
temperature. When the thermal coefficient $\omega=1$ we recover the
result for pure cluster state. At the critical point
$\omega_{c}=\sqrt{2}-1$ the state is
separable.}\label{fig:2_thermal}
\end{figure}

We already have everything we need to compute the scaling of
entanglement with temperature of a thermal 2-qubit cluster state
given by Eq.(\ref{eq:2_thermal_cluster}). The only thing that
remains to determine is a suitable candidate for the closest
separable state. The most obvious choice would be
$\sigma_{2}(\omega_{c})=\frac{1}{4}(I+\omega_{c}\sigma^{x}\otimes\sigma^{z})(I+\omega_{c}\sigma^{z}\otimes\sigma^{x})$
where $\omega_{c}=\sqrt{2}-1$. In fact this gives the correct
behavior of entanglement
$E(\sigma_{2}(\omega))=\textrm{Tr}[\sigma_{2}(\omega)\log\sigma_{2}(\omega)-\sigma_{2}(\omega)\log\sigma_{2}(\omega_{c})]$
as can be seen in Fig.(\ref{fig:2_thermal}). The gradient of the
relative entropy in the neighbourhood of the state
$\sigma_{2}(\omega_{c})$ is non-negative $\frac{\partial f}{\partial
x}(0,\tau)\geq0$ as expected.

Now that we know that for 2 qubits the cluster state at critical
temperature is a closest separable state we can ask whether there
exists a more general form of the state just like in
Sec.(\ref{thermal_ent}). It turns out that this is in fact possible,
albeit with some needed modifications. We have seen that as the
temperature increases the thermal cluster state does not approach
the state $\rho_{N}^{\ast}$ from Sec.(\ref{pure}). This can be also
seen from the non-vanishing distance between the separable states
$\sigma_{N}(\omega_{c})$ and $\rho_{N}^{\ast}$, ie
$S(\sigma_{N}(\omega_{c})||\rho_{N}^{\ast})\neq0$. As the
temperature increases the distance between the cluster state
$\sigma_{N}(\omega)$ and $\sigma_{N}(\omega_{c})$ approaches zero as
expected. However this is not true for the distance between
$\sigma_{N}(\omega)$ and $\rho_{N}^{\ast}$ due to the fact that
$S(\sigma_{N}(\omega_{c})||\rho_{N}^{\ast})\neq0$.

This problem can be overcome by not allowing the parameter $\lambda$
from Eq.(\ref{eq:2_closest_general}) to take any value from the
interval $[0,1]$. Rather we require that the lower bound of the
possible values of $\lambda$ increases with temperature. We call
this bound $\lambda^{\ast}$ and require
$\lambda^{\ast}=\lambda^{\ast}(\omega)$.

\begin{figure}
\begin{center}
\includegraphics[scale=1.1]{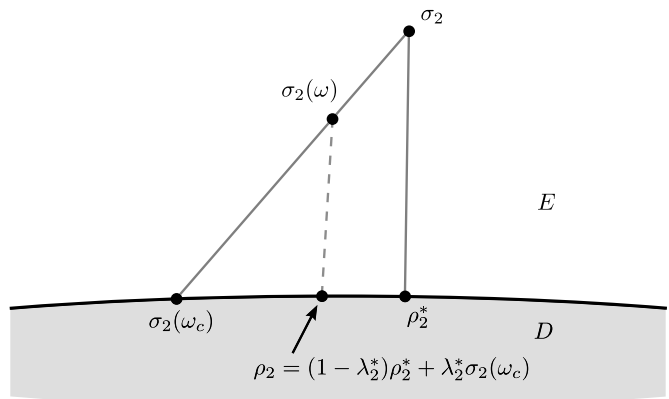}
\end{center}
\caption{Section of Hilbert space of two qubits. The shaded area $D$
represents the subspace of all separable states. $E$ designates the
set of all entangled states. As the temperature increases the set of
all separable states becomes smaller. This set is represented by all
the points found on the $E$-$D$ boundary between the states
$\sigma_{2}(\omega_{c})$ and $\rho_{2}$.}\label{fig:hilbert}
\end{figure}

As before we will first demonstrate the general principle at work on
a two-qubit cluster state $\sigma_{2}(\omega)$. First we need to
find the new parameter $\lambda^{\ast}$ by minimising the distance
between the thermal cluster state $\sigma_{2}(\omega)$ and the state
($\ref{eq:2_closest_general}$) $\rho_{2}$ for a constant
temperature. Solving
$\frac{\partial}{\partial\lambda}S(\sigma_{2}(\omega)||\rho_{2})=0$
for $\lambda$ gives the new $\lambda^{\ast}$
\begin{equation}\label{eq:2_lambda}
\lambda_{2}^{\ast}=\frac{2}{(\sqrt{2}-2)(3+\omega)}
\end{equation}
We can straight away confirm that the separable state given by this
new parameter
$\rho_{2}=(1-\lambda_{2}^{\ast})\rho_{2}^{\ast}+\lambda_{2}^{\ast}\sigma_{2}(\omega_{c})$
is also a closest separable state by calculating the gradient of
relative entropy at this state $\frac{\partial f}{\partial
x}(0,\tau)\geq0$. Therefore we can construct a new general closest
separable state
\begin{equation}\label{eq:2_new_gen_closest}
\rho_{2}(\omega)=(1-\lambda)\rho_{2}^{\ast}+\lambda\sigma_{2}(\omega_{c})
\end{equation}
where this time $\lambda\in[\lambda_{2}^{\ast},1]$. Relative
positions of all the above states in Hilbert space are illustrated
in Fig.(\ref{fig:hilbert}). One thing that the figure does not
capture is the fact that the distances
$S(\sigma_{2}||\rho_{2}^{\ast})$ and
$S(\sigma_{2}||\sigma_{2}(\omega_{c}))$ are the same. This is
because not all properties of Hilbert space can be pictured on a
two-dimensional drawing.

This method of finding closest separable states can be extended to
general $N$-qubit cluster states. The crucial part is to find the
critical temperature at which the cluster becomes fully separable.
Due to presence of bound entanglement for $N\geq3$ this is a
non-trivial task and at present it is unclear how it can be
achieved. \cite{Cavalcanti:2009} calculates upper and lower bounds
on the value of the critical temperature for cluster states of
different dimensions. However all that this changes is that the
separable state $\sigma(\omega_{c})$ would be shifted to the left of
the state $\sigma_{2}(\omega_{c})$ on the $E$-$D$ boundary in
Fig.(\ref{fig:hilbert}) since the critical temperature for $N\geq3$
is higher for than the one calculated here.

\section{Conclusion}
We have demonstrated a systematic way of constructing closest
separable state to $N$-qubit cluster states. Our method reproduces
known results for pure states and also allows us to quantify
entanglement for mixed cluster states.

For pure states the method relies on writing the density matrix of
the state in the "mixed" basis and setting any off-diagonal elements
to zero. Operationally we can construct this state from $N/2$ copies
of 2-qubit closest separable state and joining them together by
applying control-phase gates as demonstrated in Sec.(\ref{pure}).
Our method applies to cluster states of all dimensions. This is
particularly useful since cluster states of dimension $d\geq2$ are
universal resource state for quantum computation.

It also turns out that state $\rho_{N}^{\ast}$ is not the only
closest separable state. Thermal cluster state at critical
temperature $\sigma_{N}(\omega_{c})$ is also another closest
separable state to pure cluster state. Therefore any convex mixture
of these two states also minimizes the relative entropy. However at
the moment it is difficult to determine the precise temperature at
which the cluster becomes fully separable for cases of more than 2
qubits. This is due to the presence of bound entanglement.

Mixed states require an even more careful approach. When quantifying
entanglement in thermal cluster states a convex mixture of
$\rho_{N}^{\ast}$ and $\sigma_{N}(\omega_{c})$ does not minimize the
relative entropy anymore. Parameter $\lambda$ that determines the
mixture of these two states does not take any value from interval
$[0,1]$. Instead as the temperature increases the lower bound of
this interval increases as well. So the range of values the
parameter $\lambda$ can take is constricted to
$\lambda\in[\lambda^{\ast},1]$. The new lower bound $\lambda^{\ast}$
is a function of temperature and as temperature approaches the
critical value $\lambda^{\ast}\rightarrow1$. In other words both the
thermal cluster $\sigma_{\omega}$ and the closest separable state
$\rho$ approach the same state.

\ack We would like to thank Daniel Cavalcanti and Mark Williamson
for useful discussions. MH and VV acknowledge financial support from
EPSRC. VV would also like to thank Royal Society and the Wolfson
Trust in the UK, National Research Foundation and  Ministry of
Education in Singapore as well as European Union for financial
support.


\section*{References}
\begin{thebibliography}{20}
\bibitem{EPR:1935} Einstein A, Podolsky B and Rosen N 1935, {\it Phys.
Rev.} {\bf 47} 777
\bibitem{Walther:2005} Walther P, Resch K J, Rudolph T, Schenck E, Weinfurter H, Vedral V, Aspelmeyer M and
Zeilinger A 2005, {\it Nature} {\bf 434} 169
\bibitem{Vandersypen:2001} Vandersypen L M K, Steffen M, Breyta M G, Yannoni C S, Sherwood M S and
Chuang I L 2001, {\it Nature} {\bf 414} 883
\bibitem{Deutsch:1985} Deutsch D 1985, {\it Proc. R. Soc.} {\bf A
400} 73
\bibitem{Freedman:2003} Freedman M, Kitaev A, Larsen M and Wang Z
2003, {\it Bull. Amer. Math. Soc.} {\bf 40} 31
\bibitem{Raussendorf:2001} Raussendorf R and Briegel H 2001, {\it Phys. Rev.
Lett} {\bf 86} 5188
\bibitem{Hein:2004} Hein M, Eisert J and Briegel H 2004, {\it Phys.
Rev.} {\bf A 69} 062311
\bibitem{Briegel:2000} Briegel H and Raussendorf R 2000, {\it Phys. Rev.
Lett.} {\bf 86} 910
\bibitem{Biggerstaff:2009} Biggerstaff D N, Rudolph T, Kaltenbaek R,
Hamel D, Weihs G and Resch K J 2009, {\it Preprint
quant-ph/0909.2843}
\bibitem{Markham:2007} Markham D, Miyake A and Virmani A 2007, {\it New J.
Phys.} {\bf 9} 194
\bibitem{Horodecki:2009} Horodecki R, Horodecki P, Horodecki M and
Horodecki K 2009, {\it Rev. Mod. Phys.} {\bf 81}, 865
\bibitem{Raussendorf:2005} Raussendorf R, Bravyi S and Harrington J
2005, {\it Phys. Rev.} {\bf A 71} 062313
\bibitem{Wei:2004} Wei T-C, Ericsson M, Goldbart P M and Munro W J
2004, {\it Quant. Inf. Comp.} {\bf 4} 252
\bibitem{Vedral:2004} Vedral V 2004, {\it New J. Phys.} {\bf 6} 102
\bibitem{Ishizaka:2003} Ishizaka S 2003 {\it Phys. Rev.} {\bf A 67},
060301
\bibitem{Miranowicz:2008} Miranowicz A and Ishizaka S 2008, {\it Preprint quant-ph/0805.3134}
\bibitem{Vedral:1997} Vedral V 1997, {\it Phys. Rev. Lett.} {\bf 78}
2275
\bibitem{Vedral:1998} Vedral V and Plenio M 1998, {\it Phys. Rev.} {\bf A
57} 1619
\bibitem{Vedral_thesis:1998} Vedral V 1998, {\it Ph.D. thesis} Imperial
College of Science,Technology and Medicine
\bibitem{Markham:2008} Markham D, Anders J, Vedral V and Murao M
2008, {\it EPL} {\bf 81} 40006
\bibitem{Peres:1995} Peres A 1995, {\it Phys. Rev.} {\bf A 202} 16
\bibitem{Horodecki:1996} Horodecki M, Horodecki P and Horodecki R
1996, {\it Phys. Rev.} {\bf A 223} 1
\bibitem{Kay:2006} Kay A, Pachos J, D\"{u}r W and Briegel H 2006, {\it New J.
Phys.} {\bf 8} 147
\bibitem{Cavalcanti:2009} Cavalcanti D, Aolita A, Ferraro A,
Garc\'{i}a-Saez A and Ac\'{i}n A 2009, {\it Preprint
quant-ph/0909.5609}
\bibitem{Barrett:2008} Barrett S, Bartlett S, Doherty A, Jennings D
and Rudolph T 2008, {\it Preprint quant-ph/0807.4797}
\bibitem{Barnett:2009} Barnett S M 2009, \emph{Quantum Information}. Oxford University Press, Oxford
\end{thebibliography}
\end{document}